\documentclass{aa}
\usepackage{times,epsf,psfig}
\begin{document}

   \thesaurus{06
             (09.10.1;
               08.06.2;
               08.03.4;
               13.09.4)}
   \title{Giant protostellar outflows revealed by infrared imaging}

   \author{Thomas Stanke
\thanks{Visiting Astronomer, German-Spanish Astronomical Centre, 
Calar Alto
operated by the Max-Planck-Institute for Astronomy, 
Heidelberg, jointly with the Spanish National Commission for Astronomy}
\and Mark J. McCaughrean
\and Hans Zinnecker}

   \offprints{T. Stanke (tstanke@aip.de)}

   \institute{Astrophysikalisches Institut Potsdam (AIP),
              An der Sternwarte 16, D-14482 Potsdam, Germany
             }

   \date{Received; accepted}

   \maketitle

   \begin{abstract}

We present new infrared data from a survey for embedded protostellar jets
in the \object{Orion\,A} cloud. This survey makes use of the \hbox{S(1) 
$\rm v = 1$--0} 
line of molecular hydrogen at $\lambda = 2.12$\,$\mu$m to search for 
infrared jets deep inside the cloud and thus hidden from view at optical 
wavelengths. We present data on the three flows associated with the 
Herbig-Haro objects \object{HH\,43/38} \& \object{HH\,64}, \object{HH\,65}, 
and the \object{L1641-N} giant flow. \object{HH\,43}, \object{HH\,38} and 
\object{HH\,64} are part of one
H$_2$ flow extending over at least 1.4\,pc. We identify a deeply embedded 
1.3\,mm and IRAS source (\object{HH\,43\,MMS1} = \object{IRAS\,05355$-$0709C}) 
as the likely driving source, while the infrared source previously assumed 
to drive \object{HH\,43/38} (\object{HH\,43-IRS1} = 
\object{IRAS\,05357$-$0710}) is seen to drive a smaller jet. The morphology 
of \object{HH\,43-IRS1}
suggests that it is a star+disk system seen close to edge-on.
We identify another large H$_2$ flow apparently comprising the 
\object{L1641-S3} CO outflow and the redshifted lobe of the 
\hbox{\object{L1641-S}} CO outflow containing \object{HH\,65}. 
This flow extends
over at least 2.6\,pc and appears strongly curved. It is driven by 
\object{L1641-S3\,IRS}, a deeply embedded 1.3\,mm and IRAS source 
(\object{L1641-S3\,MMS1} = \object{IRAS\,05375$-$0731}).
Finally, we have found some additional large H$_2$ features to the 
east of \object{V\,380\,Ori} and the \object{HH\,1/2} system, which probably 
outline another part of the \object{L1641-N} outflow. The molecular flow
\object{MB\,20/21}, which extends to the south from \object{V\,380\,Ori}, also 
appears to be a part of the \object{L1641-N} outflow. 
      \keywords{ISM: jets and outflows -- Stars: formation -- circumstellar matter -- Infrared: ISM: lines and bands}
   \end{abstract}

%

\section{Introduction}

Star formation is known to be intimately connected with outflow activity,
especially when the star is in its main accretion phase, still deeply 
embedded in its parent cloud (e.g.\ Bontemps et al.\ 1996). 
Many low-mass young stellar objects are found
to drive well-collimated jets. Interestingly, the length-scales of these
jets have increased with advances in detector technology, namely bigger
imaging arrays. It now seems that flows from low-mass stars 
extending over several parsecs are no great exception (e.g.\ Bally 
\& Devine 1994, 1997; Devine et al.\ 1997; Eisl\"offel \& Mundt 1997; 
Reipurth et al.\ 1997, 1998; Stanke et al.\ 1999);
however, the great majority of them have been found at optical wavelengths.

In this paper we demonstrate that infrared imaging is also capable of 
revealing large scale flows. This is
complementary progress, since infrared imaging is a much more efficient way
of finding jets deep in molecular clouds and interacting with the
clouds, as these are hidden from view at optical wavelengths
(e.g.\ the infrared jets \object{HH\,211}; McCaughrean et al.\ 1994; and 
\object{HH\,212}; Zinnecker et al.\ 1998).
Only a combination of infrared and optical imaging can fully reveal the
true number of jets in a cloud and thus enable us to address questions
concerning the role of outflows from young stars for the evolution of a cloud,
e.g.\ in terms of momentum-injection to stabilize the cloud against collapse.

The data presented here are part of an infrared survey for embedded jets 
in the Orion\,A cloud. The full survey will cover an area of about 1 square 
degree with a sensitivity comparable to that of previous infrared observations
of individual protostellar jets. Initial results from the survey were
published by Stanke et al.\ (1998, Paper~I in the following). The new results 
presented here confirm that infrared imaging is an efficient way to 
search large areas for embedded flows hidden from view at optical wavelengths,
but which are bright in H$_2$ emission in the near-infrared.
 
\begin{figure*}
	\vspace{0cm}
	\vspace{0cm}
	\caption{2.12\,$\mu$m image (not continuum subtracted) of the 
	\object{HH\,43/38} region. The image size is 
	$\sim$14\,arcmin$\times$10\,arcmin (1.8\,pc$\times$1.3\,pc).
	H$_2$ features are labeled with their SMZ numbers. The cross
	marks the position of \object{HH\,43\,MMS1} 
	($\alpha = 5^h 37^m 57.\!^s5$,
	$\delta = -7\degr 07^\prime 00^{\prime\prime}$ (J2000)), which 
	apparently drives the jet
	outlined by \object{SMZ\,7-5}, 7-6, 7-7, 7-9, 7-10, and 7-14. A 
	certain symmetry of the H$_2$ features is seen around MMS1, which 
	however breaks down at larger distances: \object{HH\,43} and 
	\object{HH\,38} are without similar brightness 
	counterparts on the other side of the driving source. The apparent
	symmetry and morphology in the knots led us to suspect that the
	driving source was located
	somewhere between \object{SMZ\,7-6} and 7-7, as confirmed by the 
	subsequent discovery of MMS1. \object{HH\,43-IRS1} is 
	unrelated to the large flow, but is seen to drive a smaller jet
	as indicated by \object{SMZ\,7-8}, a bow-shock like H$_2$ feature 
	that opens back towards \object{HH\,43-IRS1}, not MMS1.}
	\label{hh43l}
\end{figure*}

\begin{figure}
	\vspace{0cm}
	\vspace{0cm}
	\caption{2.12\,$\mu$m image (not continuum subtracted) of the area 
	around \object{HH\,43-IRS1} (\object{IRAS\,05357$-$0710}). 
	\object{SMZ\,7-8} consists of a bright knot with a faint
	diffuse tail extending south-east towards \object{HH\,43-IRS1}. 
	\object{HH\,43-IRS1} itself is resolved into two stellar sources 
	plus a bipolar continuum nebula intersected 
	by a dark lane. The fainter star, located within the
	dark lane, appears to be surrounded by an almost edge on
	disklike structure causing the dark lane, and a diffuse envelope
	causing the bipolar reflection nebulosity.
	\object{SMZ\,7-8} seems to mark a bow shock caused by a jet from the 
	star+disk system. The line connecting the star and \object{SMZ\,7-8} 
	is nearly perpendicular to the dark lane, 
	suggesting that the jet is roughly perpendicular to the disk plane.
	}
	\label{hh43irs}
\end{figure}

\begin{table}
\caption{Positions of the H$_2$ features discussed in the text. The features
are located in survey fields 6, 7, and 9.}
\begin{center}
\begin{tabular}{r@{}lccccl}
\hline
\multicolumn{2}{c}{Name} & RA & Dec & \multicolumn{2}{c}{d$^a$} & Note\\
 & &\multicolumn{2}{c}{J2000.0} & arcmin & pc & \\ \hline
SMZ\,&6-2    & 5 36 30.6 & $-$6 40 00 &18.1&2.4&         \\
     &6-4A   & 5 36 32.0 & $-$6 42 28 &20.5&2.7&         \\
     &6-4B   & 5 36 33.6 & $-$6 44 12 &22.3&2.9&         \\
     &6-4C   & 5 36 35.2 & $-$6 45 40 &23.7&3.1&         \\
     &6-16   & 5 36 33.1 & $-$6 53 25 &31.3&4.1&\vspace*{1mm}\\
     &7-5    & 5 37 47.0 & $-$7 05 22 & 3.0&0.39& HH\,64   \\
     &7-6    & 5 37 51.5 & $-$7 06 08 & 1.5&0.19&         \\
     &7-7    & 5 38 01.0 & $-$7 07 39 & 1.1&0.15&         \\
     &7-8    & 5 38 03.9 & $-$7 07 47 & 1.1&0.15&         \\
     &7-9    & 5 38 05.5 & $-$7 08 30 & 2.5&0.32& HH\,43X  \\
     &7-10$^b$&5 38 10.8 & $-$7 09 25 &4.2 &0.55 &HH\,43   \\
     &7-14   & 5 38 21.8 & $-$7 11 36 &7.6&1.0& HH\,38\vspace*{1mm}\\
     &9-3A   & 5 40 25.8 & $-$7 22 14 &11.1&1.4&         \\
     &9-3B   & 5 40 23.4 & $-$7 22 48 &10.2&1.3&         \\
     &9-4A   & 5 40 43.0 & $-$7 23 29 &13.7&1.8&         \\
     &9-5A   & 5 40 23.9 & $-$7 24 29 &9.2&1.2&         \\
     &9-6A   & 5 40 15.3 & $-$7 24 25 &7.6&1.0&         \\
     &9-11A  & 5 39 53.3 & $-$7 30 57 &0.74&0.1&         \\
     &9-12A  & 5 39 45.8 & $-$7 34 44 &5.0&0.65&         \\
     &9-12B  & 5 39 42.2 & $-$7 32 51 &4.1&0.53&         \\
     &9-13A  & 5 39 49.9 & $-$7 33 54 &3.7&0.48&         \\
     &9-14A  & 5 39 40.6 & $-$7 35 26 &6.2&0.8&         \\
     &9-15   & 5 39 46.3 & $-$7 36 52 &6.9&0.9&         \\
\end{tabular}\\

\end{center}

$^a$ Separation from the driving source measured in arcmin and in pc (projected
separation assuming a distance to the Orion star forming region of 450~pc)\\ 
$^b$ Knot B2, see Schwartz et al.\ 1988\\

\end{table}

\section{Observations}

The infrared data presented here were obtained during several observing runs
(Jan.\ 10-11 1998, Oct.\ 23-26 1998, Dec.\ 5 1998)
using the 
Omega Prime wide-field near-IR camera (Bizenberger et al.\ 1998;
McCaughrean et al., in prep.) on 
the Calar Alto 3.5\,m telescope. The camera uses a 1024$\times$1024 pixel 
HgCdTe array: at 0.4\,arcsec per pixel, the field-of-view is 
6.7\,arcmin$\times$6.7\,arcmin. Images were taken through a 1\,\% filter 
centred on the \hbox{$\rm v=1$--0\,S(1)} 
line of H$_2$ at 2.12\,$\mu$m and a broad-band K$^{\prime}$ 
filter (1.944--2.292\,$\mu$m) in order to discriminate line and continuum 
sources. The images presented here are excerpts of larger mosaics, each of them
a combination of 36 overlapping images. In the 2.12\,$\mu$m filter, the net 
integration time was 10\,min in the central part of each mosaic and 5\,min 
at its edges; at K$^{\prime}$, the corresponding integration times were 1\,min and 
30\,sec. We reach a 5$\sigma$ limiting H$_2$ surface brightness of 
$\sim 10^{-18}$\,W\,m$^{-2}$\,arcsec$^{-2}$
(K$^{\prime}$$\sim$17 for continuum point sources). 
Standard data reduction techniques were used to sky subtract, flat field, 
and mosaic the data (McCaughrean et al.\ 1994).

The data discussed in this paper are part of a much larger survey in Orion\,A
split into a number of overlapping fields.
We only assign designations to those H$_2$ features which are part of the
outflows discussed here. We adopt a nomenclature similar to that
used in Paper~I, with an additional digit indicating the survey
field. For example, \object{SMZ\,7-10} means H$_2$ feature number 10 in 
survey field 7.
A complete catalogue of all features found during our survey is deferred
to a future paper. 

Millimetre continuum maps at 1.3\,mm were taken of several of the most 
interesting new H$_2$ sources during an observing run 
in February 1999, using the MPIfR 37 channel bolometer array on the IRAM
30\,m telescope on Pico Veleta. The details of these observations as 
well as the complete results
will be discussed elsewhere; here we will only present data 
related to the flows discussed in this paper.

\section{The \object{HH\,43/38} outflow}

\subsection{Previous studies}

\object{HH\,43} and \object{HH\,38} were discovered via optical photography
by Haro (1953; see Herbig 1974). CCD images of these objects were presented by 
Strom et al.\ (1986) and Eisl\"offel \& Mundt (1997), the latter discovering
another feature, \object{HH\,43X}, north-west of \object{HH\,43}. 
Infrared studies of H$_2$ line emission from \object{HH\,43} have been
reported in several papers (Schwartz et al.\ 1987, 1988; 
Zinnecker et al.\ 1989; Gredel 1994; Schwartz \& Greene 1999).
An optically invisible infrared source (\object{HH\,43-IRS1}, 
\object{IRAS\,05357$-$0710}) north-west of \object{HH\,43} was considered 
as the source of the \object{HH\,43/38} system 
(Cohen \& Schwartz 1983, 1987; Reipurth \& Aspin 1997). Cohen et al.\ 
(1984, 1985) presented far-infrared (47-130\,$\mu$m)
measurements of this source, showing it to be a cool low-luminosity object. 
At near-infrared wavelengths, the source was resolved into a double
star with some associated nebulosity (Gredel 1994; Moneti \& Reipurth 1995).
Casali (1995) measured the K-band polarisation of the source as 3.78\,\%
at a position angle of 46\,degrees. 

Searching for a molecular outflow associated with the HH objects, 
Edwards \& Snell (1983, 1984) found no
evidence for well-ordered supersonic molecular motion there, and while
Morgan \& Bally (1991) observed some evidence for high-velocity
material at the position of \object{IRAS\,05357$-$0710}, they later 
regarded this not to be a molecular outflow in a follow-up paper 
(Morgan et al.\ 1991).

\object{HH\,64} was found by Reipurth \& Graham (1988). Cohen (1990) found
a very red IRAS source (\object{IRAS\,05355$-$0709C}) close to this HH-object,
which they assumed to be its driving source.
The region was searched for CS (1--0) emission
by Tatematsu et al.\ (1993) and Morata et al.\ (1997). Tatematsu et al.\
found two cores which appear to be at the positions of the IRAS sources:
their core \#98 corresponds to \object{IRAS\,05357$-$0710}, and core 
\#97 to \object{IRAS\,05355$-$0709C}\@.
Anglada et al.\ (1989) observed an ammonia condensation at the position of the
latter core and suggested that the source responsible for the excitation
of \object{HH\,64} as well as \object{HH\,43/38} might be embedded in this 
condensation. Below we show that this hypothesis was correct.


\begin{figure*}
	\vspace{0cm}
	\vspace{0cm}
	\caption{HIRES-processed IRAS images of the \object{HH\,43} region
	(pixel size is 15\,arcsec, field size is 
	17.5\,arcmin$\times$13.75\,arcmin). 
	The positions of the HH-objects 38, 43, and 64
	(from left to right) are indicated as squares, crosses
	mark the positions of the IRAS sources 05357$-$0710 
	(\object{HH\,43-IRS1}, previously assumed to be the driving source 
	of the \object{HH\,43/38} flow; see Cohen \& Schwartz 1983, 1987)
	and 05355$-$0709C. In addition, the position of \object{HH\,43\,MMS1}
	is marked by a diamond. The positional accuracy for 
	\object{IRAS\,05355$-$0709C} is probably limited by the 
	deconvolution process and of the order of 
	$\sim$2\,pixels, corresponding to about 30\,arcsec. Within
	these errors, \object{HH\,43\,MMS1} and \object{IRAS\,05355$-$0709C}
	are coincident.}
	\label{hh43IRAS}
\end{figure*}

\subsection{New observational results}

Fig.\ \ref{hh43l} shows the region around \object{HH\,43/38} and to their 
north-west as we have imaged it.
A long chain of H$_2$ features is clearly identified, delineating a well 
collimated jet. \object{HH\,38} (\object{SMZ\,7-14}), \object{HH\,43} 
(\object{SMZ\,7-10}), and \object{HH\,43X} (\object{SMZ\,7-9})
are seen to be part of the 
south-eastern lobe of this jet. A gap is evident between features 
\object{SMZ\,7-7} and \object{SMZ\,7-6}. The latter features, along with
\object{HH\,43X} and \object{HH\,64}, are seen to form two pairs
lying roughly symmetrically around the middle of this gap. No further 
H$_2$ features (besides possibly some very faint traces of emission) are 
found to the north-west of \object{HH\,64} out to a distance of 
$\sim$8\,arcmin from the
middle of the gap, although it should be noted that the outermost part of 
the image is from data taken under rather bad weather conditions, 
leading to a nominally poorer detection limit; nevertheless, features 
with a H$_2$ brightness comparable to \object{HH\,43} can safely be excluded.
We do not detect any continuum emission in the gap between \object{SMZ\,7-6}
and 7-7 to a limiting magnitude of about K=17.5 (3$\sigma$ for a point source).

To the north-west of \object{HH\,43X} we identify a cometary H$_2$ feature
(\object{SMZ\,7-8}, see Fig.\ \ref{hh43irs}) with
a bright head to the north-west and a tail to the south-east, which is 
clearly displaced from the axis of the large flow. The tail points back
to \object{HH\,43-IRS1}, which is resolved into the components
previously reported by Gredel (1994) and Moneti \& Reipurth (1995). The
diffuse nebulosity consists of two roughly symmetric lobes separated by
a dark lane, which is oriented at an angle of $\sim$45\,degrees east of north.
The fainter star is located slightly displaced from the centre of symmetry
of the nebulosity ($\sim$0.6--0.9\,arcsec to the north-west). The dark lane
is about 2.5\,arcsec wide and at least 4\,arcsec long, corresponding to about 
1100\,AU and 1800\,AU respectively at a distance of 450\,pc. 

\begin{figure*}
	\vspace{0cm}
	\hspace{0cm}
	\hfill\parbox[b]{5.5cm}{\caption[]{\label{hh43mms} This image shows 
	a contour plot of a  1.3\,mm map of the region around 
	the \object{HH\,43/38} outflow source (resolution $\sim$15\,arcsec), 
	superimposed on a narrow band 
	2.12\,$\mu$m image. The contours start at a level of 20\,mJy and 
	increase in steps of 20\,mJy. Positions are in B1950.}}
	\vspace{1cm}
\end{figure*}

We examined HIRES-processed IRAS maps of the area around \object{HH\,43}, 
as shown in Fig.\ \ref{hh43IRAS}. \object{HH\,43-IRS1} 
(\object{IRAS\,05357$-$0710}) is detected at all four
IRAS wavelengths. Cohen (1990) found an additional source 
(\object{IRAS\,05355$-$0709C}) on the coadded IRAS images, which is not in
the Point Source Catalog, but is clearly seen in the
60\,$\mu$m and 100\,$\mu$m HIRES maps. From the 60\,$\mu$m map (with 10 
iterations
of HIRES processing) we find this source to be located at $\alpha = 5^h 37^m 
57.\!^s8$, $\delta = -7\degr 07^\prime 03^{\prime\prime}$ (J2000), which is
about 30\,arcsec north-east of the position given by Cohen (1990). However, 
the positional accuracy (as defined by the deconvolution) is probably not 
better than a few pixels,
that is about 30\,arcsec . \object{IRAS\,05355$-$0709C} is fainter than 
\object{IRAS\,05357$-$0710} at
60\,$\mu$m, but brighter at 100\,$\mu$m, consistent with the fluxes 
given for these sources by Cohen (1990) and Cohen \& Schwartz (1987).

In addition, a strong compact 1.3\,mm source was found at $\alpha = 5^h 37^m 
57.\!^s5$, $\delta = -7\degr 07^\prime 00^{\prime\prime}$ (J2000; positional
accuracy $\sim$3\,arcsec) with a flux of about
600\,mJy in an 11\,arcsec beam. This source is henceforth referred to as 
\object{HH\,43\,MMS1}. A 4\,arcmin$\times$4.6\,arcmin section of the 
1.3\,mm map is shown in Fig.\ \ref{hh43mms} as a contour plot overlaid 
on the corresponding 2.12\,$\mu$m infrared image. Some 32\,arcsec south 
and 20\,arcsec west of \object{HH\,43\,MMS1} there is an east-west 
elongated ridge of emission, which we label \object{HH\,43\,MMS2}. 
Surrounding both features there appears to be substantial extended emission
out to about 30\,arcsec north-east and 100\,arcsec to the west
and south-west of \object{HH\,43\,MMS1}. No significant emission was 
detected from \object{HH\,43-IRS1}, a result which is consistent with 
our detection limit of $\sim$50\,mJy and the 1.3\,mm flux of 47$\pm$14\,mJy
measured by Reipurth et al.\ (1993). The position of \object{HH\,43\,MMS1}
coincides with that of \object{IRAS\,05355$-$0709C} to within
a few arcseconds, and thus it is very likely the same source that is seen at
60\,$\mu$m, 100\,$\mu$m, and 1.3\,mm.

\subsection{Discussion}

The new near-infrared data clearly show that the HH-objects 38, 43, and 64
are part of a single well-collimated flow. It
extends over at least 11\,arcmin in projection (from \object{HH\,38} to 
\object{HH\,64} only) equivalent to 1.4\,pc at a distance of 450\,pc. 
This projected length is probably close to the true length of the flow,
since infrared spectroscopy of \object{HH\,43} indicates that the flow 
lies close to the plane of the sky (e.g.\ Schwartz \& Greene 1999). 

The driving source of this flow appears to be the newly discovered 1.3\,mm 
object \object{HH\,43\,MMS1}, which is associated
with the cold IRAS source 05355$-$0709C, embedded in a dense NH$_3$/CS core
(Anglada et al.\ 1989; Morata et al.\ 1997; Tatematsu et al.\ 1993), and
appears to be a very good candidate for a Class\,0 protostar 
(see Fig.\ \ref{sedhh43mms}). 
It seems reasonable to assume that the flow is intrinsically fully symmetric
and extends out to similar distances on both sides of its source with a 
total length of $\sim$2\,pc (the distance from the source to \object{HH\,38}
is about 1\,pc). Indeed, some very faint traces of emission might be present 
in our low S/N data north-west of \object{HH\,64}, but this has to be 
confirmed. A similar break in symmetry is also observed in \object{HH\,212},
the prototypical symmetric H$_2$ jet (Zinnecker et al.\ 1998), where the 
outermost bow shock is only visible in one jet lobe.

Identifying \object{HH\,43\,MMS1} as the driving source 
also removes the difficulties in explaining the morphology of some features 
(see e.g.\ Eisl\"offel \& Mundt 1997): the 
south-eastern part of \object{HH\,43} as well as the bow-shock like 
\object{HH\,43X} clearly do not point back towards \object{HH\,43-IRS1}, 
but towards \object{HH\,43\,MMS1}. 
Thus {\sl \object{HH\,43-IRS1} seems to be unrelated to \object{HH\,43} 
and the associated infrared jet}. \object{HH\,43\,MMS1} is obviously deeply
embedded, since it is only seen at the
longest IRAS and millimetre wavelengths and is completely invisible in our
K-band images. 
It is likely a Class\,0 type object, as is suggested by the model SED (Fig.\ 
\ref{sedhh43mms}), which yields a ratio of 
$L_{\rm bol}/L_{\rm submm}$ of only 11 (note that the high-frequency fit
in the two component curve in Fig.\ \ref{sedhh43mms} is only constrained by
upper limits, the derived $L_{\rm bol}/L_{\rm submm}$ of 11 thus is in fact
an upper limit). \object{HH\,43\,MMS1} thus easily fullfills the
criterion for a classification as a Class\,0 object 
($L_{\rm bol}/L_{\rm submm} < 200$) suggested by
Andr\'{e} et al.\ (1993). Additional (sub)millimetre continuum observations
are needed to further constrain the SED of \object{HH\,43\,MMS1} and to 
estimate its evolutionary stage. 

\begin{figure}
	\vspace{0cm}
	\vspace{0cm}
	\caption{Spectral energy distribution of the driving source of the
	\object{HH\,43} jet. 1.3\,mm, K-, and I-band photometry/upper 
	limits are derived from our own data, and IRAS photometry is taken 
	from Cohen (1990). In addition to the data points we show the
	curve which was used to derive the total luminosity $L_{\rm bol}$
	and the submillimetre luminosity $L_{\rm submm}$ of the source
	based on ``bluebody'' curves (see Dent et al.\ 1998) with 
	$F_\nu \propto B_\nu(T)\times(1-e^{-\tau_\nu})$ and
	$\tau_\nu \propto (\nu/\nu_0)^\beta$.
	We did not attempt to derive the circumstellar dust mass or dust 
	properties: the curves were only used to get an estimate
	of $L_{\rm bol}$/$L_{\rm submm}$ by integrating
	over $F_\nu$ from $\lambda = 3$\,mm to 300\,$\mu$m ($L_{\rm submm}$)
	and to 10\,$\mu$m ($L_{\rm bol}$). According to the criterion for a
	classification as a Class\,0 source 
	($L_{\rm bol}/L_{\rm submm} < 200$) 
	suggested by Andr\'{e} et al.\ (1993), \object{HH\,43\,MMS1}
	qualifies as Class\,0 protostar.}
	\label{sedhh43mms}
\end{figure}

Identifying \object{HH\,43\,MMS1} as the driving source 
also removes the difficulties in explaining the morphology of some features 
(see e.g.\ Eisl\"offel \& Mundt 1997): the 
south-eastern part of \object{HH\,43} as well as the bow-shock like 
\object{HH\,43X} clearly do not point back towards \object{HH\,43-IRS1}, 
but towards \object{HH\,43\,MMS1}. 
Thus {\sl \object{HH\,43-IRS1} seems to be unrelated to \object{HH\,43} 
and the associated infrared jet}. \object{HH\,43\,MMS1} is obviously deeply
embedded, since it is only seen at the
longest IRAS and millimetre wavelengths and is completely invisible in our
K-band images. 
It is likely a Class\,0 type object, as is suggested by the model SED (Fig.\ 
\ref{sedhh43mms}), which yields a ratio of 
$L_{\rm bol}/L_{\rm submm}$ of only 11 (note that the high-frequency fit
in the two component curve in Fig.\ \ref{sedhh43mms} is only constrained by
upper limits, the derived $L_{\rm bol}/L_{\rm submm}$ of 11 thus is in fact
an upper limit). \object{HH\,43\,MMS1} thus easily fullfills the
criterion for a classification as a Class\,0 object 
($L_{\rm bol}/L_{\rm submm} < 200$) suggested by
Andr\'{e} et al.\ (1993). Additional (sub)millimetre continuum observations
are needed to further constrain the SED of \object{HH\,43\,MMS1} and to 
estimate its evolutionary stage. 

The K-band morphology of \object{HH\,43-IRS1} suggests that
the southern star is surrounded by a disk-like structure or a flattened
envelope which is seen close to edge on, and a tenuous, more spherically
symmetric envelope. The star appears within the
dark lane, located close to its north-western edge. 
This suggests that the north-western part of the disk is slightly
tilted towards the
observer. The disk is seen at a position angle of $\sim$45 degrees,
in excellent agreement with the polarization angle derived by Casali (1995), 
assuming that light scattered in a tenuous envelope above a disk-like
dust configuration will yield a polarization angle
parallel to the disk plane (e.g.\ Els\"asser \& Staude 1978; 
Fischer et al.\ 1994, 1996).
Since the disk is seen close to edge-on, the star suffers substantial
extinction, which leads to its disappearance at shorter wavelengths
(see Moneti \& Reipurth 1995). The star to the north, which is the brighter
component of the system in the K-band, also disappears at shorter wavelengths
and is thus heavily extincted. It is probably located behind the star+disk
system; whether it is physically related to the system is not clear. We
cannot exclude the possibility that it is a background star, but since 
there is a dense cloud core associated with \object{HH\,43-IRS1}, it might
well be another young star embedded in the same core.
The dust lane separating the lobes of the reflection nebula
in \object{HH\,43-IRS1} seems to be much bigger than those observed in 
YSOs in the Taurus star forming region (Padgett et al.\ 1999) and the 
\object{Orion Nebula} silhouette disks
(McCaughrean \& O'Dell 1996; McCaughrean et al.\ 1998).

The connecting line between \object{SMZ\,7-8} to the north-west and the 
star in the dark lane is at 85\,degrees with respect to the dark lane, 
consistent with the usual finding that jets are perpendicular to the 
disks of their driving sources. The morphology of \object{SMZ\,7-8} is
suggestive of a bow shock like structure heading away from the disk+star 
system. We thus regard the disk+star system as
the source of a second jet which includes knot \object{SMZ\,7-8}
(see Fig.\ \ref{hh43irs}), independent of the large 
\object{HH\,38/43/64} flow from \object{HH\,43\,MMS1}. There is no conclusive
evidence for a counter jet in this system, although if we reflect the 
north-western bow symmetrically about the disk, a putative counter-jet
would lie at the northern edge of \object{HH\,43}, making it hard to see
against this bright feature. Furthermore, the
suggested inclination of the disk also implies that the north-western
part of the jet is the blueshifted lobe, whereas the counter-lobe would
be redshifted, running into the cloud, and thus possibly heavily obscured.  

\begin{figure*}
	\vspace{0cm}
	\vspace{0cm}
	\caption{2.12\,$\mu$m image (not continuum subtracted) of the 
	\object{L1641-S} and \object{L1641-S3} region. The image size is 
	$\sim$17\,arcmin$\times$17\,arcmin ($\sim$ 2.2\,pc$\times$2.2\,pc). 
	The contours show the distribution of 
	high-velocity CO (adapted from Morgan et al.\ 1991). 
	Solid lines indicate redshifted, dotted lines 
	blueshifted gas. The dashed lines mark the boundaries of the 
	maps presented by Morgan et al.; the large eastern square surrounds
	the area of flow \object{MB\,40}, the smaller rectangular piece to 
	the west the area of flow \object{MB\,41}. The position of 
	\object{IRAS\,05380$-$0728} is marked by a white cross; 
	\object{L1641-S3\,IRS} is coincident with
	\object{IRAS\,05375$-$0731} within the errors.}
	\label{re50l}
\end{figure*}

\begin{figure*}
	\vspace{0cm}
	\vspace{0cm}
	\caption{Continuum subtracted image of the same area as in 
	Fig.\ \ref{re50l}. H$_2$ features related to the suggested flow 
	from \hbox{\object{L1641-S3\,IRS}} = \object{L1641-S3\,MMS1} 
	(as discussed in the body of the paper) are labeled with their SMZ
	numbers. In addition, some more H$_2$ features are found around 
	\object{Re\,50\,N} and at the northern edge
	of the image. The cross marks the position of \object{L1641-S3\,IRS}.}
	\label{re50pure}
\end{figure*}

\begin{figure*}
	\vspace{0cm}
	\hspace{0cm}
	\hfill\parbox[b]{5.5cm}{\caption[]{\label{l1641s3mms} This image
 	shows a 1.3\,mm contour plot of the region around 
	the \object{L1641-S3} outflow source, superimposed on our narrow band 
	2.12\,$\mu$m image. The contours start at a level of 25\,mJy and 
	increase in steps of 30\,mJy. The lower left and the upper right corner
	were not covered by our mm-mapping. The features in the lower
	right corner are noise peaks. Positions are in B1950.}}
	\vspace{1cm}
\end{figure*}

\section{The \object{L1641-S} and \object{L1641-S3} flow}

\subsection{Previous studies}

\object{IRAS\,05380$-$0728} is located about 10\,arcmin to the south and 
34\,arcmin to the east of the \object{HH\,43} region and is one of the more 
luminous young stellar objects in the \object{L1641} cloud 
($\sim$250--370\,L$_{\sun}$; Reipurth \& Bally 1986; Cohen 1990), 
probably a low- to intermediate-mass star which may have undergone 
an FU Orionis type outburst (Strom \& Strom 1993).
Fig.\ \ref{re50l} shows a 2.12$\mu$m image of the area
around this source and its associated reflection nebulosities \object{Re\,50}
and \object{Re\,50\,N} (Reipurth 1985; Strom et al.\ 1989; Heyer et al.\ 1990;
Casali 1991; Chen \& Tokunaga 1994; Hodapp 1994; Casali 1995; Colom\'{e} et
al.\ 1996; Reipurth \& Aspin 1997). Reipurth \& Bally (1986) and Fukui 
et al.\ (1986) independently discovered
a bipolar high velocity molecular gas flow apparently from 
\object{IRAS\,05380$-$0728}, the so-called \object{L1641-S} outflow. 
Further mapping of the high velocity gas in this flow presented by 
Morgan et al.\ (1991, flow \object{MB\,40}) showed
that the northern redshifted lobe of the flow describes a long curve to the 
north and west of the IRAS source position (see contour-overlay on the 
infrared image in Fig.\ \ref{re50l}). \object{HH\,65} is located 
within this lobe (Reipurth \& Graham 1988). 
Far-infrared and submillimetre measurements for \object{IRAS\,05380$-$0728} 
have been presented by Reipurth et al.\ (1993), Colom\'{e} et al.\
(1996), Chini et al.\ (1997), Zavagno et al.\ (1997), and Di Francesco 
et al.\ (1998); see also Cohen (1990). Morgan et al.\ (1990) found two 
radio continuum sources in the region surrounding \object{Re\,50\,N}; one 
is close to the IRAS and 2\,$\mu$m source position of the central source 
of the \object{Re\,50\,N} nebulosity, while the second is displaced by 
50\,arcsec to the west. There is no NH$_3$ core associated with this 
second source (Wouterloot et al.\ 1988, 1989).

Another IRAS source (\object{IRAS\,05375$-$0731}) is found about 3\,arcmin
south and 7.9\,arcmin west of \object{IRAS\,05380$-$0728}. It is associated 
with the molecular outflow \object{L1641-S3} (Fukui 1988; Fukui et al.\ 1989).
The distribution of high velocity molecular material 
around \object{IRAS\,05375$-$0731} has been mapped by Wilking et al.\ 
(1990) and Morgan et al.\ (1991, flow \object{MB\,41}, see overlay of 
CO contours on the infrared image in Fig.\ \ref{re50l}). While Morgan 
et al.\ (1991) found a bipolar distribution,
Wilking et al.\ (1990) found a superposition of blue- and redshifted gas,
possibly indicating a flow lying close to the plane of the sky.
Near-infrared images (Chen \& Tokunaga 1994; Hodapp 1994; Strom et al.\ 
1989) revealed a small group of rather faint stars close to the IRAS source 
position; it is not entirely clear if any of them are the direct counterpart 
to the IRAS source, but the reddest one was labeled as \object{L1641-S3\,IRS}
by Chen \& Tokunaga. Also associated with this region are a radio continuum 
source (Morgan et al.\ 1990) and a water maser (Wouterloot \& Walmsley 1986),
although the latter was not redetected in a later study by Felli 
et al.\ (1992). The source has been detected by Price et al.\ (1983) as 
\object{FIRSSE\,101} in the far-infrared; further measurements in the 
far-infrared/submillimetre range are 
presented by Zavagno et al.\ (1997). Finally, there is an NH$_3$ core
associated with \object{IRAS\,05375$-$0731} (Wouterloot et al.\ 1988, 1989)

\subsection{New observational results}

Fig.\ \ref{re50l} shows a 17\,arcmin$\times$17\,arcmin 2.12\,$\mu$m narrow 
band image of the \object{L1641-S}/\object{L1641-S3} area, while 
Fig.\ \ref{re50pure} shows the same region after continuum subtraction. 
The main feature is a large system of H$_2$ knots and filaments to the 
north of \object{Re\,50} labelled \object{SMZ\,9-4}, \object{SMZ\,9-5}, and 
\object{SMZ\,9-6}. \object{HH\,65} is seen to be part of the much bigger 
feature \object{SMZ\,9-6}, including 
a bright, smoothly curving filamentary structure \object{SMZ\,9-6A} to the 
north-west of \object{HH\,65}. Further east, a number of fainter, more 
diffuse features (together labeled \object{SMZ\,9-5}) define a curving path which first heads eastwards, then
turns slightly to the south, and then more to the north again. 
Finally it ends in a multiple filamentary structure (\object{SMZ\,9-4}),
with the filaments curving
in a similar way to the filament \object{SMZ\,9-6A}. 
In the south-west corner of
the image we find several faint, diffuse H$_2$ structures (\object{SMZ\,9-11},
\object{SMZ\,9-12}, \object{SMZ\,9-13}, \object{SMZ\,9-14}, and 
\object{SMZ\,9-15}). Further faint structures
are also visible to the west and south of \object{Re\,50\,N}. Finally, a 
few more H$_2$ features (\object{SMZ\,9-3}) are found north-east and 
north-west of the brightest H$_2$ filament. The near-infrared source 
designated as \object{L1641-S3\,IRS} appears extended in our images and 
is located at $\alpha=5^h 39^m 55.\!\!^s 8$,
$\delta = -7\degr 30^\prime 28^{\prime\prime}$ (J2000).

Fig.\ \ref{l1641s3mms} shows an overlay of a contour plot of a 1.3\,mm
map of the \object{L1641-S3} outflow source region on the 2.12\,$\mu$m image.
A strong pointlike 1.3\,mm source is found at $\alpha=5^h 39^m 55.\!\!^s 9$,
$\delta = -7\degr 30^\prime 28^{\prime\prime}$ (J2000), coincident
with \object{L1641-S3\,IRS} within the positional errors of the
1.3\,mm maps ($\sim$ 3\,arcsec). A second fainter millimetre continuum
source is found 20\,arcsec to the west and 55\,arcsec to the north. Further 
north, at about $\alpha = 5^h 39^m 57^s$, $\delta = -7\degr 26^\prime 
50^{\prime\prime}$ (J2000), we find a large patch of extended emission
(size about 4\,arcmin$\times$2\,arcmin). There appears to be some low-level
extended emission 4\,arcmin west of \object{L1641-S3\,MMS1} as well.

\subsection{Discussion}

As in the case of the \object{HH\,43} flow system, our new sensitive wide field
infrared data suggest a revised picture for the \hbox{\object{L1641-S}} and 
\object{L1641-S3} region. The dominating outflow seems not to be driven by 
the luminous source associated with \object{Re\,50\,N}, but rather by 
\object{L1641-S3\,IRS} (\object{IRAS\,05375$-$0731}).
This flow includes the CO outflows
\object{L1641-S3} and the redshifted lobe of \object{L1641-S}, and is traced
by infrared H$_2$ emission over much of its length.

The H$_2$ features \object{SMZ\,9-11} to 9-15 in the blueshifted lobe of 
\object{L1641-S3} (which has not been mapped in CO), suggest a wide
opening angle ($\sim$40\,degrees) of the flow on this side. The redshifted
lobe is defined by the redshifted molecular gas of the \object{L1641-S3} 
and \object{L1641-S} flows (\object{MB\,41} and \object{MB\,40}). As is seen
in Fig.\ \ref{re50l}, these CO lobes appear to trace a single outflow lobe 
when placed alongside each other. 
The flow shows very strong bending. Close to the source, it is oriented
at a position angle of about 60\,degrees, then turns north to a position angle
of about 30\,degrees, then at about the location of \object{HH\,65} and 
the bright H$_2$ filament \object{SMZ\,9-6A} (and also the peak in the 
redshifted CO emission), it bends strongly to an eastward flow direction 
at \object{SMZ\,9-5}, before finally turning back north to roughly its 
original position angle of 60\,degrees at \object{SMZ\,9-4}. This behaviour 
is reminiscent of the wiggling of the \object{HH\,46/47} jet (e.g.\
Heathcote et al.\ 1996), albeit on a much larger scale; for example, the 
bright curving H$_2$ filaments in \object{SMZ\,9-6} mimic the H$\alpha$ 
wisps found in \object{HH\,46/47} at the positions of apparent bends in the 
jet. Heathcote et al.\ suggested that the gas does not really flow along the 
channel defined by the Herbig-Haro emission, but ballistically along the 
original direction, with the wiggles reflecting changes in the direction of 
ejection of the gas at the source. This is probably the case here as well. 
The direction of the flow at its end and close to the source appear to be the
same, thus it might be delineating one period, assuming that the changes are
periodic, e.g.\ caused by a companion star.
Assuming a typical tangential flow velocity of
about 200\,km/s, the distance from the source to the outermost part of the
flow (1.8\,pc) translates to a period of about 9000\,yr. 
If we assume that the wiggle of the jet is simply caused by periodic changes of
the source position in a binary orbit (see e.g.\ Fendt \& Zinnecker
1998), this period would imply a separation of the binary components
of order 1000\,AU or a few arcsec at the distance to Orion. However, the
apparent widening of both outflow lobes with distance from the source
and the large amplitude of the wiggles cannot be explained by simply
shifting around the outflow source in any reasonable binary orbit, and 
rather suggest a
precessing outflow than just a moving source. A precessing
jet would be caused by a precessing disk, which would be expected in a
binary system with non-coplanar disk and orbital planes. Following
the arguments given by Terquem et al.\ (1999), a precession period of 
9000\,years implies a binary separation (very roughly) on the order 10 to 
100\,AU, corresponding to angular separations of the order 0.1 arcsec in Orion.
It should be possible to resolve separations of a few arcsec or a few tenths
of an arcsec with existing or future instruments in the mid-infrared
and millimetre wavelength ranges, allowing a test of the above
assumptions. 

Alternatively, we may simply see a poorly collimated outflow,
possibly also responsible for exciting the faint features to the west of
\object{Re\,50\,N} and \object{SMZ\,9-3}, with the brighter H$_2$ features 
only imitating a bending flow, or illuminating
particular parts of a large outflow cavity. In this respect it is also 
interesting to note that the luminosity of the driving source 
\object{L1641-S3\,MMS1} is rather high ($\sim 70 L\sun$, compared to
$\sim 5 L\sun$ in \object{HH\,43\,MMS1}), possibly suggesting that 
\object{L1641-S3\,MMS1} may be an intermediate-mass young stellar object.
The \object{L1641-S3} outflow may thus give further support to the 
suggestion that flows from intermediate- and high-mass protostars 
might be systematically less well collimated than
those from their low-mass counterparts, as is seen e.g. in \object{DR\,21}
(Davis \& Smith 1996), \object{G192.16-3.82} (Shepherd et al.\ 1998), the 
molecular outflow \object{G75\,C} in the \object{ON2} core (Shepherd et al.\
1997), and 
\object{HH\,288} (McCaughrean \& Dent, in prep.). Note however that 
Davis et al.\ (1998) find that flows from high mass protostars may 
nevertheless be driven by collimated jets like their low mass counterparts.

\begin{figure}
	\vspace{0cm}
	\vspace{0cm}
	\caption{Same as Fig.\ \ref{sedhh43mms} for 
	\object{L1641-S3\,MMS1}. 1.3\,mm, K-, and I-band photometry/upper 
	limits are derived from own data, IRAS photometry is taken from the
	point source catalog, and additional submillimetre photometry is 
	taken from Zavagno et al.\ (1997).
	With $L_{\rm bol}/L_{\rm submm} \sim 80$--90, 
	\object{L1641-S3\,MMS1} also qualifies as Class\,0 protostar.}
	\label{sedl1641s3mms}
\end{figure}

The driving source of the outflow appears to be \object{L1641-S3\,IRS} = 
\object{L1641-S3\,MMS1}.
The spectral energy distribution of this source is suggestive of a Class\,0
type object ($L_{\rm bol}/L_{\rm submm} \sim 80$--90, see 
Fig.\ \ref{sedl1641s3mms}),
although its detection at the shorter IRAS wavelengths and its association
with a small near-infrared nebulosity suggests that it may be approaching
the Class\,I stage.

\begin{figure}
\vspace{0cm}
\vspace{0cm}
	\caption{2.12\,$\mu$m image of the southern lobe of the 
	\object{L1641-N} outflow centered roughly on $\alpha=5^h 36^m 26^s$,
	$\delta = -6\degr 37^\prime 40^{\prime\prime}$ (J2000). The
	image size is $\sim$13\,arcmin$\times$33\,arcmin, corresponding to
	$\sim$1.7\,pc$\times$4.3\,pc. H$_2$ features related to that flow are
	marked, as well as some other well known objects in the area.}
	\label{l1641n}
\end{figure}

\section{The \object{L1641-N} outflow}

In Paper~I, we found evidence for a
large scale H$_2$ jet from the embedded \object{L1641-N} cluster, with the more
prominent features (\object{SMZ\,23}) extending to the south of the cluster.
Meanwhile, a chain of optically visible Herbig-Haro objects 
(\object{HH\,306}-\object{HH\,310}) has
been found to the north of the cluster by Reipurth et al.\ (1998; see
also Mader et al.\ 1999), which is probably the 
counterpart to the infrared H$_2$ flow. However, the northern Herbig-Haro 
flow terminates at a distance of 6.3\,pc north of the cluster, whereas 
the southern infrared flow as delineated mainly by \object{SMZ\,23} was seen to
extend only $\sim$1.8\,pc south of the cluster.

In Fig.\ \ref{l1641n} we present a 2.12\,$\mu$m image of the region south
of the \object{L1641-N} cluster with H$_2$ features marked. We find a number
of large H$_2$ features to the north-west, west, and south-west of 
\object{V\,380\,Ori}
(\object{SMZ\,6-2}, \object{SMZ\,6-4}, and \object{SMZ\,6-16}) which appear 
to be the continuation of the southern H$_2$ flow from \object{L1641-N}, 
beyond \object{SMZ\,23}. 
Thus we suggest that the newly found features form the counterpart
to the large northern Herbig-Haro flow. The flow heads due south
at its origin in \object{L\,1641-N} and then smoothly bends to the east down 
to \object{SMZ\,6-4}.
\object{SMZ\,6-16} is then located due south of \object{SMZ\,6-4}, which 
makes its association with the flow somewhat uncertain; however, it might 
indicate a bending of the flow back to the original north-south direction. 
\object{SMZ\,6-16} has a bright rim at
its southern edge and some more diffuse emission to the north, similar to 
\object{SMZ\,26B} (\object{HH\,87/88}), which is the southern terminating
working surface in the \object{HH\,34} giant flow. Thus, \object{SMZ\,6-16}
might indicate a large bow shock 
structure in a flow running from north to south. Including the newly
discovered features, the southern infrared lobe of the \object{L1641-N}
flow is now seen to extend over almost 30\,arcmin ($\sim$4\,pc projected
length).

The region around \object{V\,380\,Ori} has been searched for high-velocity 
molecular gas by Edwards \& Snell (1984) and Levreault (1988). 
Among other features, they found a large lobe of redshifted gas extending 
from about the region around \object{V\,380\,Ori}
to the south (see also Morgan \& Bally 1990; Morgan et al.\ 1991; flows
\object{MB\,20}/\object{MB\,21}). No blueshifted counterflow was found. 
\object{V\,380\,Ori} was suggested as the driving source for this 
north-south oriented monopolar flow. In contrast, Corcoran \& Ray (1995) 
suggested an east-west oriented outflow from \object{V\,380\,Ori}.
The newly discovered H$_2$ features (including \object{SMZ\,6-16}, somewhat 
off the track of the flow as suggested by the other H$_2$ features)
appear superimposed 
on the north-south oriented redshifted CO outflow lobe found by Edwards \& 
Snell (1984) and Levreault (1988). We suggest that this CO lobe is 
not related to \object{V\,380\,Ori}, but indeed another piece of
the southern, redshifted part of the \object{L1641-N} outflow.

\section{Summary and conclusions}

Using data from our wide-field infrared imaging search for H$_2$ jets in
the \object{Orion\,A} molecular cloud, we have demonstrated
that parsec-scale jets can be found at infrared wavelengths.
In particular, the previously known features \object{HH\,43}, \object{HH\,38},
and \object{HH\,64} are found to be connected by several newly discovered
H$_2$ features into one large flow. A second large flow is found to consist
of the previously known \object{L1641-S3} CO outflow and the redshifted 
lobe of the \object{L1641-S} CO outflow containing \object{HH\,65}. Again, 
the full extent of this flow is revealed by infrared imaging. In both cases, 
we also found deeply embedded millimetre continuum sources which we suggest to 
be Class\,0 protostars driving the flows. Finally, we have shown that the 
southern lobe of the 
\object{L1641-N} flow, which is traced by H$_2$ emission, may extend over 
at least 4\,pc, i.e.\ over a similar
length as the northern lobe, which is traced by optically visible
HH-objects. We suggest that the known molecular outflow lobe 
\object{MB\,20/21} (originally thought to be driven by \object{V\,380\,Ori})
is connected to this giant flow. 

Besides the three flows discussed in this paper, other
giant flows are known in \object{Orion\,A}\@. In addition to the prototype 
\object{HH\,34} system (Bally \& Devine 1994; Eisl\"offel \& Mundt 1997;
Devine et al.\ 1997), the \object{HH\,1/2} system probably belongs to 
this class of objects (Ogura 1995). Some flows in the \object{OMC\,2/3}
region also seem to extend over large distances, including the 
\object{HH41/295} flow (Reipurth et al.\ 1997) 
and the H$_2$ flow J of Yu et al.\ (1997). Evidence for
further large-scale flows from the \object{L1641-N} cluster is reported by 
Reipurth et al.\ (1998) and Mader et al.\ (1999). The discovery of so
many giant flows in a single cloud appears to indicate that
giant flows are not exceptionally rare systems, but may indeed be more the
rule than previously thought.

The redshifted lobes of the \object{L1641-S3} and the \object{L1641-N} 
flows are both seen to be associated with substantial amounts of 
relatively slow moving molecular gas as traced by the CO emission over 
most of their length. The molecular gas has probably been entrained more or
less in situ, not at the base of the flows.
This implies that both flows plough through the dense gas
of the cloud itself, not through the rather thin medium around the cloud
like most of the optically-visible giant HH flows. Thus the flows
from these two protostars alone are seen to affect a significant volume of
the cloud. This gives additional support to the idea that flows from young
low mass stars may play an important role in the dynamical evolution
of a cloud and for the regulation of star formation, not only on small
scales (disruption of the cloud cores associated with the protostar driving
the flow), but also on large scales (injection of momentum, excitation of
turbulence, stabilisation of the cloud
against collapse). However, a more complete discussion of
that point is deferred to a future paper based on the complete H$_2$ survey
for protostellar jets in Orion.  

The sources of the two newly recognized giant flows, \object{HH\,43\,MMS1} and
\object{L1641-S3\,MMS1}, are both deeply embedded, bright at 1.3\,mm,
very red at IRAS wavelengths, and qualify as Class\,0 protostars; 
only one of them (\object{L1641-S3\,MMS1}) appears to be associated with a
small, faint near-infrared nebulosity. Similar behaviour 
is found for most of the Orion parsec-scale flows, implying that the
most pronounced large-scale Herbig-Haro and H$_2$ flows are driven
by very young objects, typically infrared Class\,I and even Class\,0 type
objects. Thus the largest and presumably most energetic outflows are
associated with the youngest objects, not the more evolved objects
like optically-visible T\,Tauri stars.

\begin{acknowledgements}
      Many thanks are due to Frank Bertoldi, Ernst Kreysa, Karl Menten, 
      Fr\'{e}d\'{e}rique Motte, Bernd Weferling, and Robert Zylka for their 
      help during the Pico Veleta observing run and first aid in reducing 
      the 1.3\,mm data. Thanks are also due to the referee, Chris Davis, for 
      his comments which helped to strengthen some points.
      This work was supported by the
      \emph{Deut\-sche For\-schungs\-ge\-mein\-schaft, DFG\/} project
      number Zi 242/9-1.
\end{acknowledgements}

\end{document}